\documentclass[aps,pre,showpacs,showkeys,amssymb,amsmath,superscriptaddress,reprint]{revtex4-1}

\usepackage{amsthm}
\usepackage{amsmath}
\usepackage{graphicx}
\usepackage{color}

\usepackage{verbatim}

\usepackage{amssymb}
\usepackage{bm}

\usepackage{hhline}

\begin{document}

\title{Biased random walks with finite mean first passage time}

\author{Christin Puthur}
\affiliation{Department of Physics, Indian Institute of Science Education and Research, Bhopal, India}

\author{Prabha Chuphal}
\affiliation{Department of Physics, Indian Institute of Science Education and Research, Bhopal, India}

\author{Snigdha Thakur}
\affiliation{Department of Physics, Indian Institute of Science Education and Research, Bhopal, India}

\author{Auditya Sharma}
\affiliation{Department of Physics, Indian Institute of Science Education and Research, Bhopal, India}

%\date{\today}

\begin{abstract}
A power-law distance-dependent biased random walk model with a tuning parameter
($\sigma$) is introduced in which finite mean first passage times are
realizable if $\sigma$ is less than a critical value $\sigma_c$. We
perform numerical simulations in $1$-dimension to obtain $\sigma_c
\sim 1.14$. The three-dimensional version of this model is related to the phenomenon of chemotaxis. Diffusiophoretic theory supplemented with
coarse-grained simulations establish the connection with the specific value of $\sigma = 2$ as a consequence of in-built solvent diffusion. A variant
of the one-dimensional power-law model is found to be applicable in the context of a stock investor devising a strategy for extricating their portfolio out of loss.
\end{abstract}

\maketitle

\section{Introduction}
\label{sec:introduction}

In a variety of stochastic phenomena, it is of interest to track when
a random process first reaches a threshold
value~\cite{redner,weiss1981first,bar1998mean,bray2013persistence}. This
is called first passage. For example, the decision to buy or sell
stock is made depending on when the fluctuating stock price reaches a
particular level for the first time. Another example is the diffusion
of a particle in the presence of an absorbing boundary such that the
diffusion stops the first time the particle reaches the absorbing
boundary~\cite{chandra}. First passage properties of random walks have
applications in various fields of science including physics~\cite{montroll1969random},
finance~\cite{jiang2008perpetual}, ecology~\cite{fauchald2003using} and chemistry~\cite{kim1958mean}.

For unbiased random walks in one-dimensional space, it can
be shown that with probability unity, first passage \emph{will} occur,
although the \emph{mean first-passage time} is infinite~\cite{firstpassagetracy}. 
It would be of interest to understand whether and in what proportion the introduction of bias
can result in a finite expectation value of the mean first-passage time.
The exploration of this question in the context of a particular
type of bias, namely power-law bias is the central objective of this
paper.  We introduce a one-parameter $(\sigma)$ power-law
distance-dependent biased random walk model in which the expectation
value of the first-passage time is found to be finite when $\sigma$ is
less than a critical value. 

Distance-dependent biased stochastic motion is in fact, a ubiquitous
phenomenon in nature. Bacterial motion is one such example where the
bacterium moves towards food or away from toxin respectively, by the
mechanism of chemotaxis~\cite{berg1993random}. Another example where
such biased random walk is observed is in the biochemical cycle of
molecular motors~\cite{Thomas2113}. This process at the simplest level
can be compared to a biased random walk in which the bias is dependent
on the position of the random walk at every instant. In this paper, we
make a connection between chemotaxis and the power-law biased random
walk model introduced above. To do this, we perform coarse-grained
simulations for a pair of small colloidal particles, one of which is a
source of chemical reaction and the other chemotaxes towards it by
diffusiophoresis.

If the movement of stock is well-modeled by an unbiased random walk,
the implication for an investor awaiting the price to hit a particular
value is that they can be certain this will happen, but they may have
to wait forever. Since this is an unsatisfactory situation to be in,
the possibility of controlled bias to make the expected first-passage time
finite would be greatly desirable.  A variant of the power-law bias
model which could potentially be applied in such a realistic stock
market investing scenario is therefore introduced and studied. Here,
the external bias comes naturally from the buying of stock
intermittently, and the number of stock purchased is tuned according
to the power-law model. The finite expectation value of first-passage
time could potentially be exploited as an active strategy by an
investor with sufficient surplus funds to extricate a loss-making
stock out of loss.

The organization of this paper is as follows. In the next section, we
review the first-passage properties of unbiased random walks, and
random walks with a constant bias. The third section introduces the power-law bias
model, and some novel simulation techniques we have developed to study it, and 
then discusses the results. The fourth section describes in some detail the chemotaxis
example supplemented with molecular simulations. The fifth section describes
the variant model specially designed as a strategy for a loss-making investor, followed
by simulation results. A concluding section wraps up the paper.

\section{Review of first passage in random walks}
\label{sec:models}

\subsection{Unbiased case}

\begin{figure}
\centering
\includegraphics[scale=0.33]{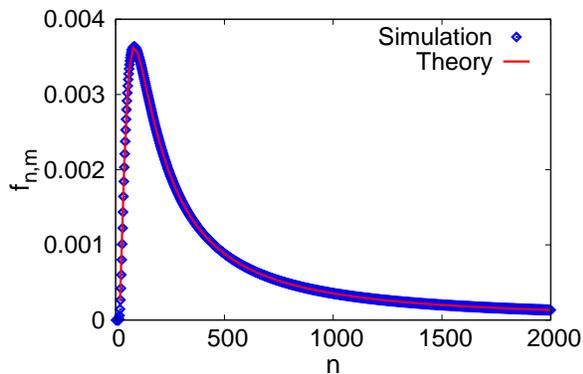}
\caption{The first-passage probability as a function of the number of
  steps taken for an unbiased random walk starting from $m=16$ to the
  origin. The diamond symbols depict the points from the
  simulation. The solid line is the curve obtained from
  Eqn.~\ref{eq:4}.}
\label{fig:fpsim}
\end{figure}

Let $p_{n,m}$ denote the probability for a one-dimensional random walk starting at some distance $m>0$ from the origin to be at the origin at the $n^\text{th}$ step.
\begin{equation}
p_{n,m}=\frac{1}{2^n}\frac{(n)!}{\left(\frac{n+m}{2}\right)!\left(\frac{n-m}{2}\right)!}=\frac{1}{2^n}\binom{n}{\frac{n+m}{2}}
\end{equation}
since the total number of possible paths in $n$ steps is $2^n$ out of which $\binom{n}{\frac{n+m}{2}}$ paths reach the origin after $n$ steps. We consider the probability for the random walk to reach the origin for the first time after $n$ steps, i.e., the first-passage probability to the origin denoted by $f_{n,m}$.

\begin{figure}
\centering
\begin{minipage}[b]{.49\textwidth}
 \includegraphics[scale=0.33]{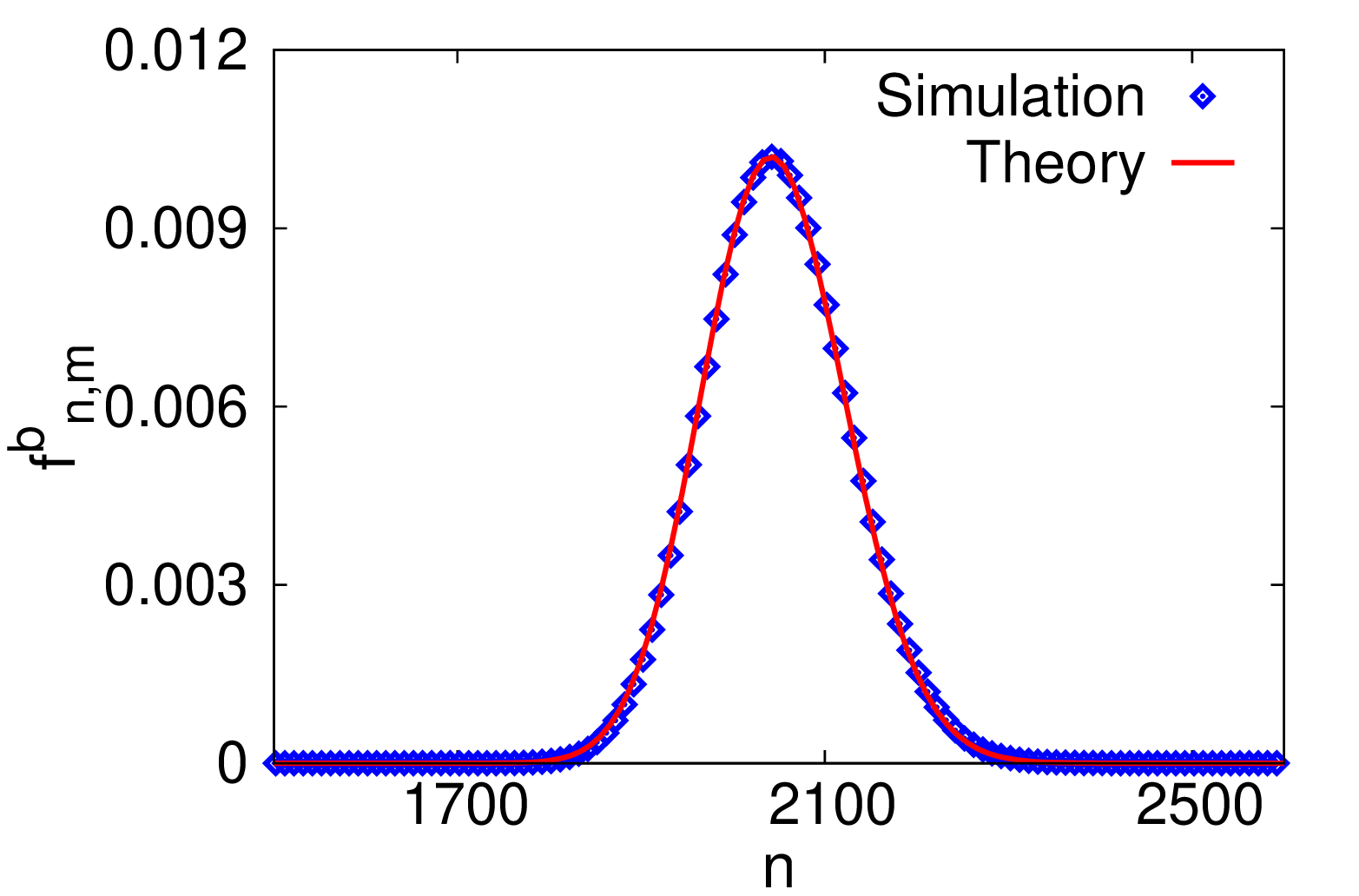}
 \caption{The first-passage probability as a function of the number of
   steps taken for a biased random walk with $p-q=0.5$ starting from
   $m=1024$ to the origin. The diamond symbols depict the points from the
   simulation. The solid line is the curve obtained from
   Eqn.~\ref{eq:7}.}
 \label{fig:bfpsim}
\end{minipage}\hfill
\begin{minipage}[b]{.49\textwidth}
 \includegraphics[scale=0.33]{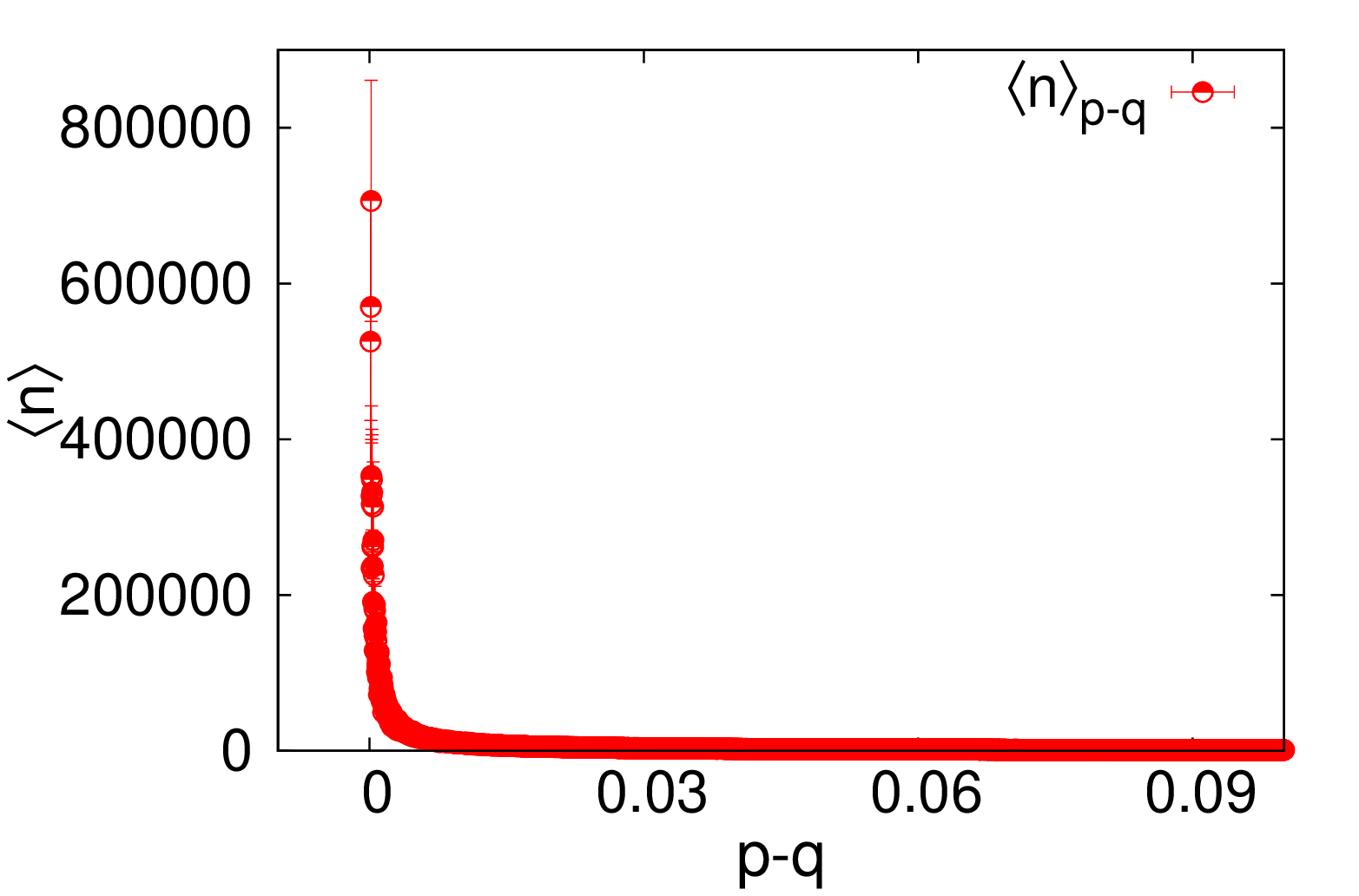}
 \caption{The plot shows the simulation results of the mean first-passage time $\langle n\rangle_{p-q}$ to the origin for a sample of 500 random walks starting from $m=100$ with a constant bias of $p-q=0$ to $p-q=0.1$ towards the origin. High error-bars for $p-q=0$ is indicative of infinite mean first-passage time.}
 \label{fig:bmfpt}
\end{minipage}
\end{figure}

Here, we assume that either $m$ and $n$ are both even or both odd,
because otherwise the random walk cannot be at the origin at the $n^\text{th}$
step. The number of paths from $m$ reaching the origin for the first
time at the $n^\text{th}$ step is $\frac{m}{n}\binom{n}{\frac{n+m}{2}}$~\cite{feller}. 
Thus, the first-passage probability to the origin for a random walk starting at $m$ is
\begin{equation}
f_{n, m}=\frac{1}{2^{n}}\frac{m}{n}\binom{n}{\frac{n+m}{2}}=\frac{m}{n}p_{n, m}.
\end{equation}
The binomial distribution of $p_{n,m}$ is approximated by the
Gaussian distribution for large $n$. We have $p_{n,m}=\frac{1}{2^n}\frac{(n)!}{\left(\frac{n+m}{2}\right)!\left(\frac{n-m}{2}\right)!}$. In
the limit of large $n$, using Stirling's formula, $\ln{(n!)}=(n+\frac{1}{2})\ln{n}-n+\frac{1}{2}\ln{2\pi}+O(n^{-1})$,
where $O(n^{-1})\rightarrow 0$ as $n\rightarrow\infty$, and the series expansion
$\ln{\left(1\pm\frac{m}{n}\right)}=\pm\frac{m}{n}-\frac{m^2}{2n^2}+O(m^3/n^3)$,
we find that when $n\rightarrow\infty$ and $m\ll n$,
$\ln{p_{n,m}}\simeq-\frac{1}{2}\ln{n}+\ln{2}-\frac{1}{2}\ln{2\pi}
-m^2/2n$. Thus, the asymptotic formula is
\begin{equation}
p_{n,m}=\left(\frac{2}{\pi n}\right)^\frac{1}{2}\exp(-m^2/2n).
\end{equation} 
So for large $n$, the first-passage probability to the origin at the $n^\text{th}$ step for a random walk starting at $m$ is given by 
\begin{equation}
f_{n,m}=\frac{m}{n}\left(\frac{2}{\pi n}\right)^\frac{1}{2}\exp(-m^2/2n).
\label{eq:4}
\end{equation}
We find that the simulation agrees with this equation for $f_{n,m}$ as seen in Fig.~\ref{fig:fpsim}.

The first-passage probability to an absorbing
point may be found by comparing the random walk to diffusion~\cite{redner}. 
In continuous space and time, the first-passage
probability to the origin (absorbing point) for a random walk starting
at $x_0$ at time $t$ is
\begin{equation}
F(0,t)=\frac{x_0}{\sqrt{4\pi Dt^3}}e^{-x_0^2/4Dt}.
\end{equation}
We can see that $\int\limits_0^\infty F(0,t)dt=1$ which means the
random walk or diffusing particle is sure to reach the absorbing point
eventually.

Although the random walk is sure to reach the origin, the mean time
for the first-passage to the origin is infinite. The average time
required by a random walk starting from $m$ to reach the origin for
the first time is its mean first-passage time to the origin. It is
given by $\langle n \rangle=\sum\limits_{n=0}^\infty nf_{n,m}$. For
the unbiased random walk, it follows from Eqn.~\ref{eq:4} that
as $n\rightarrow\infty$, $f(n,m)\sim\frac{1}{n^\frac{3}{2}}$ and the mean first-passage time,
$\langle n\rangle=\sum\limits_{n=0}^\infty nf_{n,m}\sim\sum\limits_{n=0}^\infty\frac{1}{\sqrt{n}}\rightarrow\infty$. Thus, the mean time for the
unbiased random walk to reach the origin for the first time is infinite.

\subsection{Constant bias case}

We now consider a random walk starting at $m>0$ for which the probability of taking a step towards the origin is $p$ and the probability of taking a step in the opposite (positive) direction is $q$, where $p+q=1$. Here, the constant bias or drift towards the origin is $p-q$. In order to find the first-passage probability to the origin at the $n^\text{th}$ step for such a random walk, the number of paths is the same as in the unbiased case, i.e., $\frac{m}{n}\binom{n}{\frac{n+m}{2}}$. Thus, for the biased case, the first-passage probability is given by
\begin{equation}
f^b_{n, m}=\frac{m}{n}p^{\frac{n+m}{2}}q^{\frac{n-m}{2}}\binom{n}{\frac{n+m}{2}}=\frac{m}{n}p^b_{n,m},
\end{equation}
since $\frac{n+m}{2}$ steps are taken towards the origin and
$\frac{n-m}{2}$ steps in the positive direction. For large $n$, we
approximate this binomial distribution by the Gaussian distribution
that is centred about the mean~\cite{norm}, which is now displaced from the origin. Thus,
for large $n$, the first-passage probability to the origin for such a
biased random walk starting from $m$ is
\begin{equation}
    f^b_{n,m}=\frac{m}{n}\sqrt{\frac{1}{2\pi npq}}e^{-(m-n(p-q))^2/2(4npq)}.
    \label{eq:7}
\end{equation}
In Fig.~\ref{fig:bfpsim}, the graph of $f^b_{n,m}$ obtained by simulation is compared with the curve from Eqn.~\ref{eq:7}.

When $p-q>0$, the mean
first-passage time $\langle n\rangle$ is finite but when $p=q=0.5$, it
reduces to the unbiased case and $\langle n\rangle$ is infinite. Fig.~\ref{fig:bmfpt} compiles data from a simulation. 
It is seen that for $p-q=0$, the mean first-passage
time $\langle n\rangle$ shoots up with high error bars whereas for
$p-q>0$, the error bars are small. The high error bars show that the
average time taken will not approach a limit if the sample size is
increased implying that the mean first-passage time is infinite.

\section{Power-law biased random walk}

\begin{figure}
\centering
\begin{minipage}[b]{.49\textwidth}
  \includegraphics[scale=0.33]{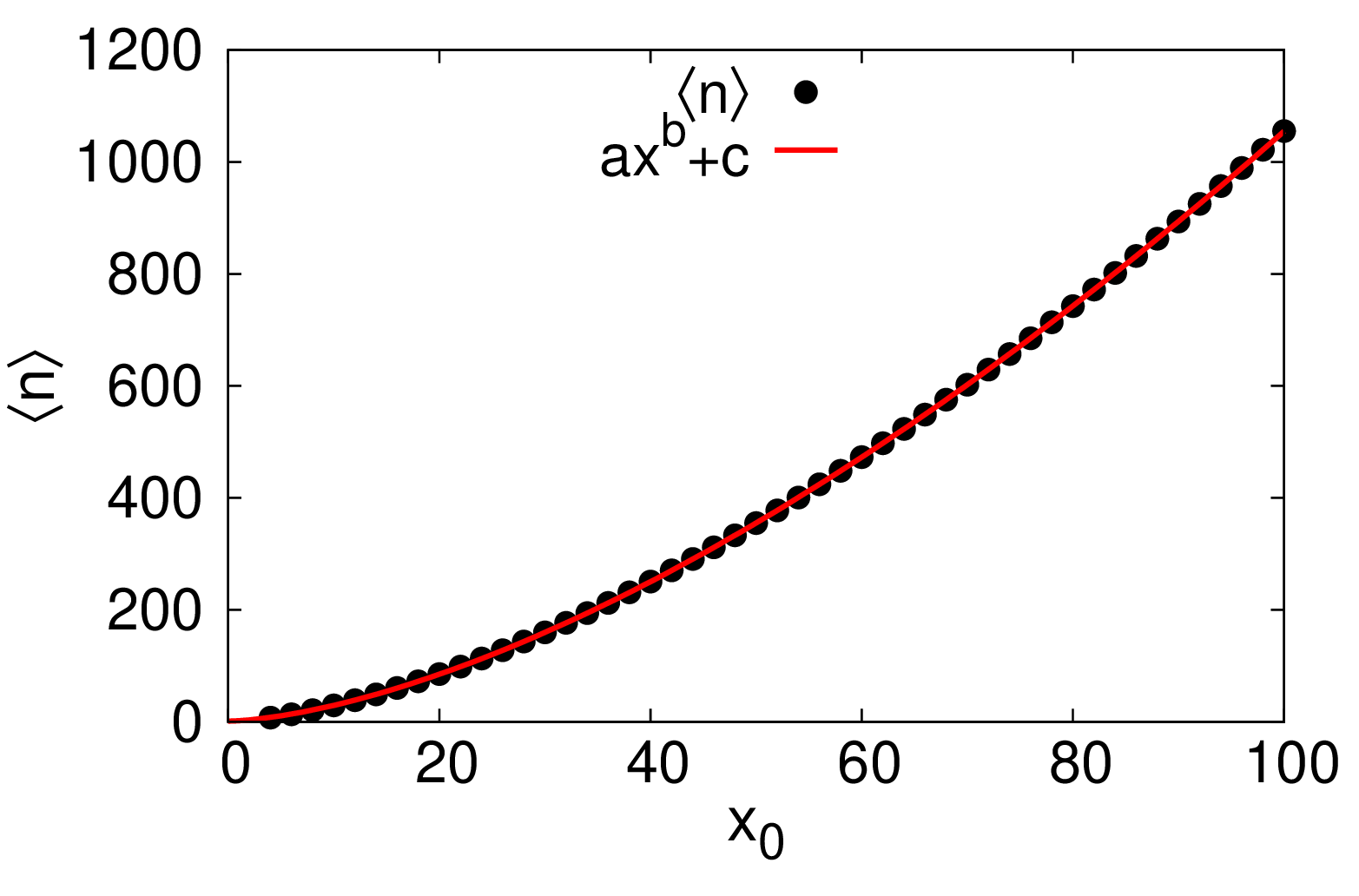}
 \caption{Variation of the mean first-passage time $\langle n\rangle$
   with the initial distance from the origin $x_0$ for $\sigma=0.6$. A
   fit for the graph with the function $ax^b+c$ gives
   $a=0.750\pm0.001$, $b=1.574\pm0.0003$ and $c=
   1.228\pm0.076$. The exponent $b$ is close to $\sigma+1$.}
 \label{fig:tvsr0}
\end{minipage}\hfill
\begin{minipage}[b]{.49\textwidth}
 \includegraphics[scale=0.33]{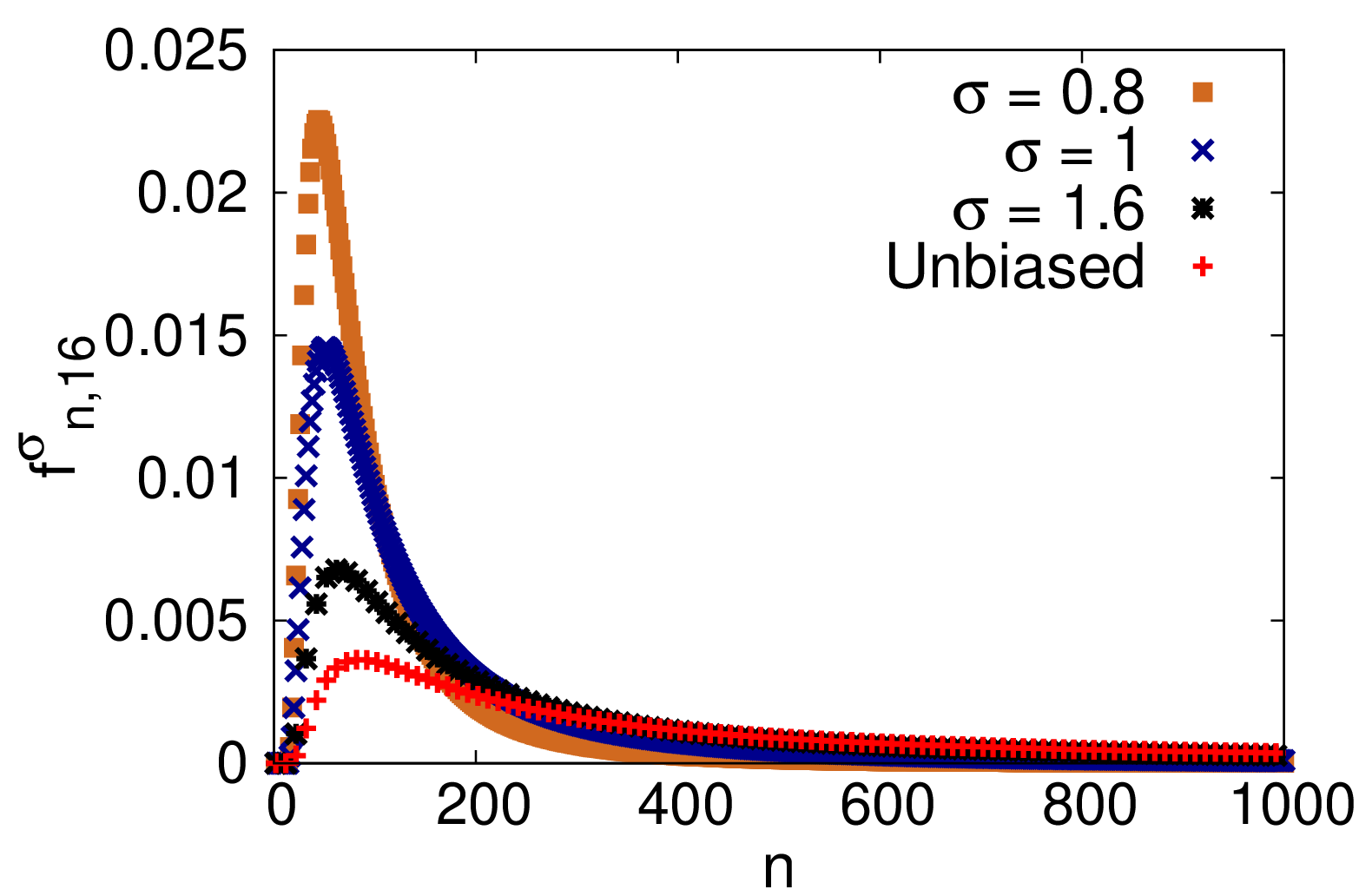}
 \caption{First-passage probability as a function of the number of steps in one dimension for different values of $\sigma$ and initial distance from the origin $x_0=16$. As $\sigma$ is decreased, we see that the large $n$ tails of the probability curves become steeper.}
 \label{fig:fpppl}
\end{minipage}
\end{figure}

\label{sec:bias}
\subsection{Model}

Although a constant bias in a random walk yields finite mean first-passage time, the cost of a constant driving force is very high and unlikely to be seen in natural systems. Here, we introduce a one-parameter $(\sigma)$ power-law distance-dependent biased random walk model to attain finite mean first-passage time. The model is inspired from chemotaxis, in which a living cell experiences a distance-dependent drift towards a target in response to a chemical gradient produced by the target \cite{chemotaxis}. We consider the origin to be the location of the target and design the bias towards it to vary with the position of the random walk $x$ as
$\frac{1}{{x}^\sigma}$ where $\sigma>0$. The average time taken to
get the fixed target captured by the diffusing cell or particle corresponds to the mean first-passage time of the random walk to the origin. We investigate how the mean first-passage time to the fixed target varies with $\sigma$.

In one-dimensional space, the biased random walk starting from $x_0>0$
operates according to the following rule. At each step, the random
walk has a probability $p=\frac{1}{2}+\frac{1}{2x^\sigma}$ to take a
step towards the origin and a probability
$q=\frac{1}{2}-\frac{1}{2x^\sigma}$ to take a step in the positive
direction away from the origin, where $x$ is the location of the
random walk at the current step. The driving force in the random walk
gives it a drift velocity $\frac{dx}{dt}=p-q=\frac{1}{x^\sigma}$
towards the origin. From a rough calculation of the mean first-passage
time to the origin: $\langle n\rangle\sim\int
tdt\sim|\int\limits_{x_0}^0x^\sigma dx|$, we deduce that the mean
first-passage time $\langle n\rangle\sim{x_0}^{\sigma+1}$. Support for this
is seen in Fig.~\ref{fig:tvsr0} for the case $\sigma=0.6$.

In this model, $\sigma$ decides the strength of the bias towards the
target and on varying $\sigma$ we find that the mean first-passage
time to the origin becomes finite below a critical value
$\sigma_c$. This is evident from our simulations in two ways. Firstly,
Fig.~\ref{fig:fpppl} shows the plots obtained for the first-passage
probability for random walks starting from $x_0=16$ for different
values of $\sigma$. As the bias is increased, the large $n$ tail of
$f^{\sigma}_{n,x_0}$ becomes steeper. We saw in the unbiased case that
$f_{n,m}\sim \frac{1}{n^\frac{3}{2}}$ as $n\rightarrow\infty$. If we
posit that for our model, $f_{n,x_0}^{\sigma}\sim\frac{1}{n^\alpha}$
for large $n$, the mean first-passage time $\langle
n\rangle\sim\sum\limits_{n=0}^\infty\frac{1}{n^{\alpha-1}}$. This
series is convergent if $\alpha-1>1$ and divergent if $\alpha-1\leq
1$. Thus, if the bias is strong enough, i.e., if $\sigma$ is
sufficiently small, we may have $\alpha>2$ and finite mean
first-passage time. In other words, there occurs a critical value
$\sigma_c$ below which the weighted average of the probability curve
becomes finite. Secondly, in Fig.~\ref{fig:mfptpl} which shows results
of simulations of an ensemble of random walks, large error-bars in the
mean first-passage time after a certain value of $\sigma$ indicates a
transition from finite to infinite mean first-passage time in the
sense that the error-bars would not shrink if the sample size is
increased~\cite{firstpassagetracy}. This transition seems to occur
near $\sigma_c\approx 1.1$ for random walks starting from
$x_0=100$. In the unbiased case, the large $n$ behaviour of
first-passage probability is independent of the starting distance
$x_0$. Therefore, we expect that the critical value $\sigma_c$ is also
independent of $x_0$.

\subsection{Simulation}

\begin{figure}
\centering
\begin{minipage}[b]{.49\textwidth}
  \includegraphics[scale=0.33]{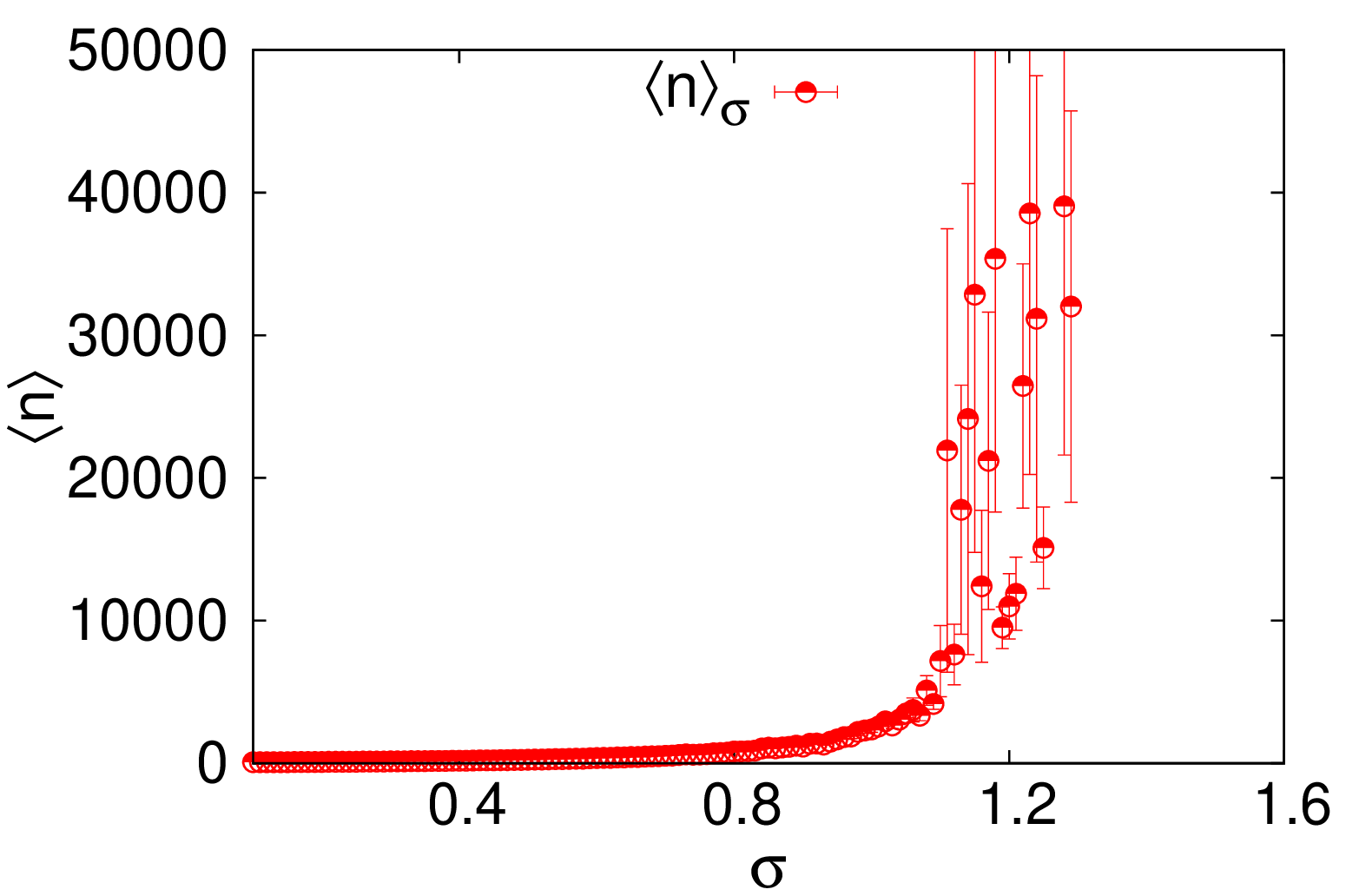}
 \caption{The plot shows the simulation results of the mean first-passage time $\langle n\rangle_{\sigma}$ to the origin for a sample of 1000 random walks starting from $x_0=100$ for different values of $\sigma$. High error-bars after a critical value of $\sigma$ is indicative of infinite mean first-passage time.}
 \label{fig:mfptpl}
\end{minipage}\hfill
\begin{minipage}[b]{.49\textwidth}
 \includegraphics[scale=0.33]{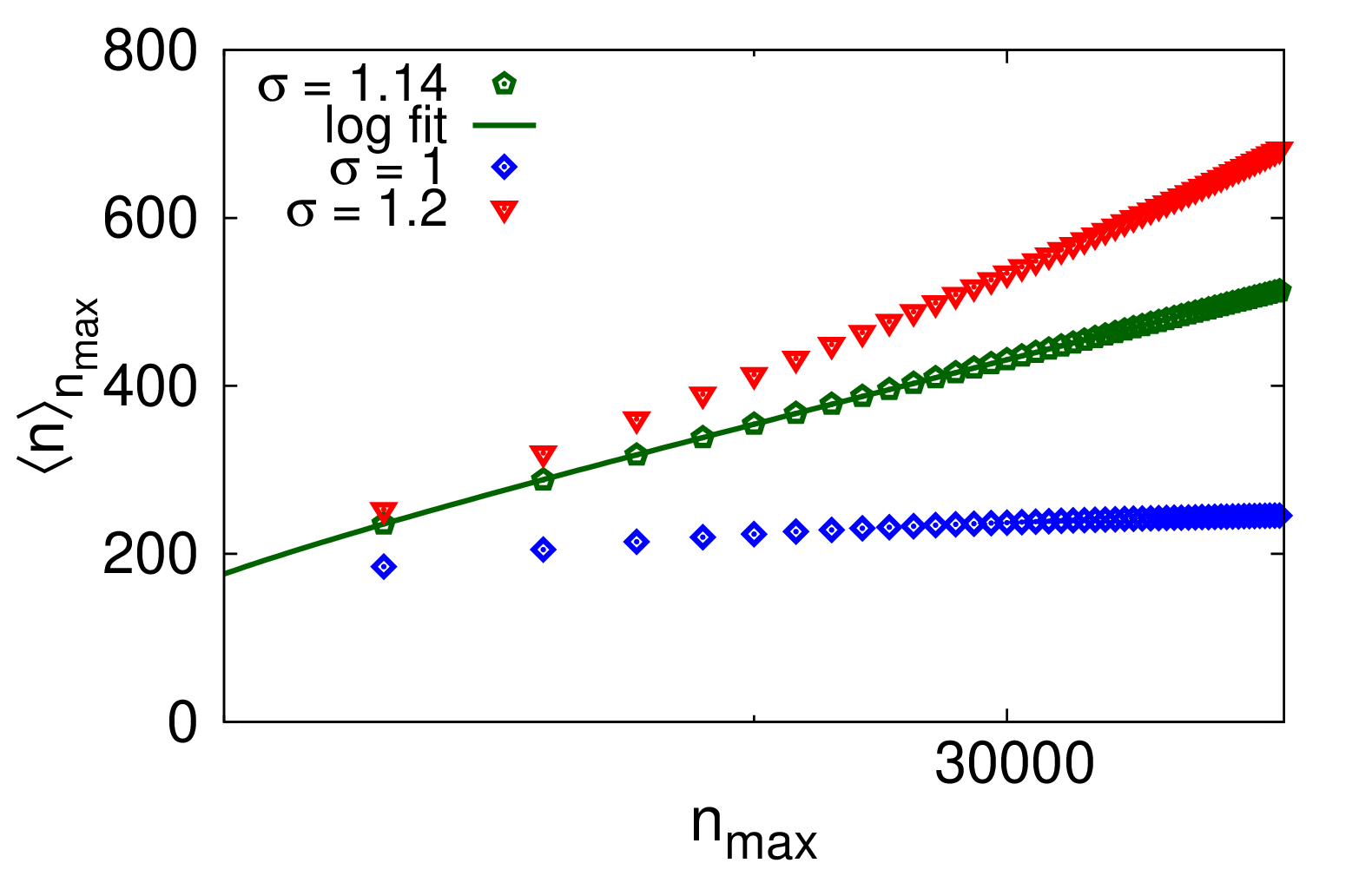}
\caption{$\langle n \rangle_{n_{\text{max}}}=\sum\limits_{n=0}^{n_{\text{max}}} nf_{n,x_0}$ as a function of $n_{\text{max}}$ for different values of $\sigma$ for an initial distance from the origin $x_0=16$. We find that the logarithmic fit works well for $\sigma=1.14$ indicating $\sigma_c\approx1.14$.}
\label{fig:critical}
\end{minipage}
\end{figure}

Here we describe the simulation technique used to obtain the curves of
first-passage probability with respect to the number of steps as shown
in Fig.~\ref{fig:fpsim}, Fig.~\ref{fig:bfpsim} and
Fig.~\ref{fig:fpppl}. As a concrete example, consider an unbiased one-dimensional random walk
starting from $x_0=4$. To obtain the first-passage probability to
the origin as a function of the number of steps, we take a
one-dimensional array of size $x_0+n_\text{max}$ and initialize the $4^\text{th}$
array element with $1$ and all other array elements with $0$. This
corresponds to the probability of the random walk being at $x_0=4$ to
be equal to $1$ initially. At the next step, the probability
distributes itself equally to its two nearest neighbours, since the
random walk can move to the left or to the right with equal
probability. This is continued at each step as shown in Table~\ref{table}.

%\begin{center}
\begin{table}
\begin{tabular}{|p{0.9cm}| p{0.8cm} |p{0.8cm}| p{0.8cm}| p{0.8cm}| p{0.8cm}| p{0.8cm}| p{0.8cm}| p{0.8cm}|}
\hline
  & $x=0$ & $x=1$ & $x=2$ & $x=3$ & $x=4$ & $x=5$ & $x=6$ & $x=7$\\
%\end{tabular}\\
%\begin{tabular}{||p{0.7cm}|p{0.8cm}|p{0.8cm}|p{0.8cm}|p{0.8cm}|p{0.8cm}|p{0.8cm}|p{0.8cm}|p{0.8cm}|}
\cline{1-9}
$n=0$ & $0$ & $0$ & $0$ & $0$ & $1$ & $0$ & $0$ & $0$\\\cline{2-9}
$n=1$ & $0$ & $0$ & $0$ & $0.5$ & $0$ & $0.5$ & $0$ & $0$\\\cline{2-9}
$n=2$ & $0$ & $0$ & $0.25$ & $0$ & $0.5$ & $0$ & $0.25$ & $0$\\\cline{2-9}
$n=3$ & $0$ & $0.125$ & $0$ & $0.375$ & $0$ & $0.375$ & $0$ & $0.125$\\\cline{1-9}
\end{tabular}
\caption{An illustration of the simulation technique.\label{table}}
\end{table}
%\end{center}

To obtain the first-passage probability at each step, the value
accumulated at the $0^\text{th}$ array element at each step is plotted with
respect to the number of steps taken. The $0^\text{th}$ element is reset
to zero after every step, because that fraction of an infinite number
of random walks have already reached the origin for the first time and cannot have
another first-passage to the origin. For the random walk with the
power-law bias the same simulation method can be employed. In this case, the
splitting of probability at each step will not be equal but will be
distance-dependent. For instance, at the first step, the probability
distributes as $\frac{1}{2}+\frac{1}{2x_0^\sigma}$ and
$\frac{1}{2}-\frac{1}{2x_0^\sigma}$ to the left and right neighbours
respectively. By this technique, numerically exact first-passage
probability at the $n^\text{th}$ time step can be found for different $\sigma$
but it is exact only up to finite time $n_{\text{max}}$.

We make a better estimate of $\sigma_c$ using the exact first-passage
probability curves obtained from this simulation technique. 
The mean first-passage time is given by $\langle n \rangle=\sum\limits_{n=0}^\infty nf_{n,x_0}$. Since we know the exact
value of $f_{n,x_0}$ only up to a maximum number of steps $n_{\text{max}}$,
we study the dependence of $\langle n \rangle_{n_{\text{max}}}=\sum\limits_{n=0}^{n_{\text{max}}} nf_{n,x_0}$ on $n_{\text{max}}$. This quantity should diverge for
$\sigma>\sigma_c$ and approach a constant value for
$\sigma<\sigma_c$. Since we posit that
$f^\sigma_{n,x_0}\sim\frac{1}{n^\alpha}$ in the long-time limit, at
the critical value $\sigma_c$, $\langle n \rangle_{n_{\text{max}}}=\sum\limits_{n=0}^{n_{\text{max}}}nf^{\sigma}_{n,x_0}\sim\int\limits^{n_{\text{max}}}\frac{n}{n^\alpha}dn\sim\log{n_{\text{max}}}$.

In Fig.~\ref{fig:critical}, we see that $\langle n \rangle_{n_{\text{max}}}$ appears to diverge for larger $\sigma$ and
approach a constant value for smaller $\sigma$. By curve fitting we
see that the logarithmic fit works excellently for $\sigma=1.14$ from which we deduce the critical value $\sigma_c\approx1.14$.

\section{Chemotaxis as an example of biased random walk}
\label{sec:chemotaxis}

Chemotaxis is the movement of a particle in response to a chemical
gradient present in the environment by a biased random
walk. Classically the microscopic chemotaxis was studied mostly in the
context of biological systems ~\cite{Alder, Alon, berg04}. However,
recent studies have explored this process for synthetic colloidal
particles~\cite{hong2007, deprez2017}, that has the potential
applications in the area of targeted particle delivery, microfluidics
etc. To establish a connection between the colloidal chemotaxis and
biased random walk, here we study the dynamics of a pair of small
colloidal particles using hybrid Molecular Dynamics - Multiparticle Collision
Dynamics (MD-MPCD). Among the two colloids, one is chemically active
stationary target (T), while the other is a biased random walker
(W) which moves towards the target (T) using the mechanism of
diffusiophoresis~\cite{anderson1989colloid}. It is the combination of
diffusiophoresis and thermal fluctuations that makes W exhibit
chemotaxis or biased random walk toward T.

\subsection{Walker (W) velocity using diffusiophoretic theory}

\begin{figure}[htbp]
	\centering
	\includegraphics[scale=0.15]{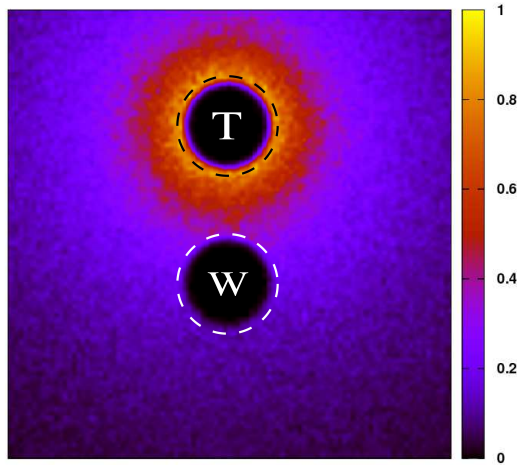}
	\caption{Formation of inhomogeneous concentration field around the the walker (W) due to the chemical reaction that converts species $A$ (fuel) to $B$ (product) in the reaction, $A + T \rightarrow B + T$ on the target (T) sphere. The walker W  moves by the diffusiophoretic mechanisms due to the asymmetry of the concentration field in its vicinity.}
	\label{fig:intro}
\end{figure}

Here, we consider a catalytically active stationary T sphere with
radius $R_1$, that converts the species $A$ (fuel) to species $B$
(product) by irreversible chemical reaction $A + T \rightarrow B + T$
on it. The W sphere with radius $R_2$ is catalytically inactive and
is separated from the target by an initial distance $R_0$ in three-dimensional
space. In such a configuration, the chemical reaction on the T
sphere, will generate a concentration gradient in the system, as shown
in Fig.~\ref{fig:intro}, which will induce a body force on W due to
different interaction potentials ($V_{W, A}\ne V_{W, B}$) of these
species with the walker. Such body force leads to the diffusiophoretic
movement of the colloidal walker with a mean velocity component along
the line of centers between the two spheres due to the axial symmetry
of the system. Below we describe the analytical method to estimate the
walker (W) velocity using diffusiophoretic theory
~\cite{anderson1986transport}.

%We consider two spheres, a catalytically active sphere  (target $T$) with radius $R_1$ and a catalytically inactive sphere (walker W) with radius $R_2$. These spheres, shown in Fig.~\ref{fig:intro}, are taken to be separated by a distance $L$ in three-dimensional space. Two solute species $A$ (fuel) and $B$ (product) take part in the irreversible chemical reaction $A + T \rightarrow B + T$ on the target sphere. While more general interactions may be considered, we further assume that the interaction potentials of these species with the target sphere are the same, $V_{1, A}=V_{1, B}$, while they are different for the walker, $V_{2, A}\ne V_{2, B}$. In this circumstance, the concentration gradient in the system arising from chemical activity on $T$ will induce  a body force on the W. The diffusiophoretic mechanism will then operate and lead to a mean velocity component along the line of centers between the two spheres due to the axial symmetry of the system. In the continuum description, our interest is in the value of the mean velocity that results from this mechanism, as well as the forms of the concentration fields that accompany it.
\par

The walker moving in a chemical gradient will experience an
inhomogeneous distribution of $B$ particles in its vicinity. The
interaction of the fuel ($A$) and product ($B$) particles with the
surface of the walker will rise to body forces, which, because of
momentum conservation will lead to a pressure and velocity gradient
within the boundary layer around the walker. As a result a fluid slip
velocity $v_s$ will be generated around the boundary layer given by
the following expression~\cite{golestanian2007designing}:
\begin{equation}
v_s=-\frac{k_{\beta}T}{\eta}(\nabla C_{B})\Lambda,
\end{equation}
where $\eta$ is the viscosity of solution, $C_{B}$ is the
concentration of product molecules on the outer edge of the boundary
layer, and $\Lambda$ estimates the strength of interaction between the
solute molecules and the surface of the colloid. The velocity of
colloid can be calculated by averaging the slip velocity over the
entire surface $\varsigma$ of the colloidal particle and can be given
as~\cite{snigdha:14, reigh:18dimerform}
\begin{equation}
V=-<v_s>_{\varsigma}.
\end{equation}

By solving the diffusion equation $D \nabla^{2}C_{A}(r)=0$ with help
of radiation boundary condition $\kappa_{0}C_{A}(R_{1},t)=\kappa_{D}
R_{1}\hat{r}.\nabla{C_{A}}(R_{1},t)$, the concentration field of fuel
and product particles can be achieved. Here $R_{1}$ is the radius at
which reaction takes place and $C_{A}(r \rightarrow \infty)=C_{0}$ for
continuous fuel supply at the boundaries to maintain the
steady-state. $C_{A}+C_{B}=C_{0}$, $D$ is the diffusion coefficient of
the solute particles, $k_{D}=4\pi R_{1}D$ is Smoluchowski rate
constant and $\kappa_{0}$ is intrinsic reaction rate constant.  By
solving the diffusion equation, concentration field of product
particles can be given as
\begin{equation}
C_{B}(r)=\frac{\kappa_{0} C_{0}}{\kappa_{0}+ \kappa_{D}} \frac{R_{1}}{r}.
\end{equation}

At any time instant $t$, let ${\bf {R}}(t)$ be the vector distance from
W to T. Then the instantaneous velocity of W sphere can be
calculated by averaging the gradient of product concentration over the
outer edge of W sphere which has radius $R_2$. By averaging the
gradient over the outer edge and projecting velocity of W sphere
along the direction of ${\bf {R}}(t)$,
$V_{z}(t)=\hat{\bf{R}}(t).{\bf{V}}(t)$~\cite{snigdha:14, kapral2007}.
\begin{equation}
V_z(t) =  \frac{2 k_{B} T C_0 \Lambda}{3 \eta} \frac{k_{0}}{k_{0} + k_{D}} \frac{R_1}{R(t)^2} \equiv \frac{\lambda}{R(t)^2}, 
\label{eq:vz}
\end{equation}
where $\Lambda =\int_{0}^{\infty} dr \: r (e^{-\beta V_{B}(r) - e^{-\beta V_{A}(r)} })$.

Further, the time evolution of separation between the spheres can be
obtained by integrating Eqn.~\ref{eq:vz}. We find $R(t) = [R_0^3 -
  3\lambda t]^{1/3}$, where $R_0$ is the initial separation between
the spheres. Thus the time taken by the walker to reach the encounter
distance $R_f = R_1+R_2$ is given by
\begin{equation}
\tau=\frac{(R_{0}^{3}-R_{f}^{3})}{3 \lambda}.
\end{equation}

\subsection{Microscopic dynamics}
The microscopic dynamics is a coarse-grain dynamics where molecular
dynamics (MD) is combined with multiparticle collision (MPC) dynamics
~\cite{footnote:mpc1}. In this scheme, fluid is modelled by $N_{sol}$ point
like particles with mass $m_s$ having the positions ${\bm{r}}_i$ and
velocities $\bm{v}_i$. The interaction between the fluid particles is
governed by multiparticle collision (MPC), whereas the interaction between 
colloids and fluid particle is through MD. MPC dynamics has two steps:
streaming and collision. In the streaming step the dynamics is evolved by
Newton's equation of motion. 
%The forces for solving Newton's equation of motion are determined by the intermolecular potentials between colloids-fluid particles and colloid-colloid particles. 
In the collision step, all the fluid particles are sorted into small cubic
cells of length $a$, which is larger than the mean free path. In
every collision step the relative velocity of fluid particles is
rotated around a randomly chosen axis by an angle $\phi$ with respect
to the center of mass velocity of every cell. The velocity of $i^\text{th}$
fluid particle after collision is ${\bm{v}_i}^{'}=\bm{v}_{cm}
+\bm{\omega}(\phi)(\bm{v}_i-{\bm v}_{cm})$, where $\bm{v_{cm}}$ is the
center-of-mass velocity of fluid particles in a cell, and $\bm\omega$ is a
rotation matrix. To ensure Galilean invariance, a random shift is
being given to the collision cells~\cite{ihle:01}. The dynamical
method described above conserves mass, momentum and energy
locally. 

The colloids (T and W) interact with fluid particles through repulsive
Lennard-Jones (LJ) potential $V=4 \epsilon [ \left( \rho/r
  \right)^{12}- \left( \rho/r \right)^6+(1/4) ] $ within the cutoff
$r_{c}=2^{(1/6)} \rho$. Here $\epsilon$ and $\rho$ are energy and
distance parameters respectively. In addition, repulsive LJ potential
is included to take care of the excluded volume interaction between the
colloids. To include the diffusiophoretic effects for W motion, we
choose the interaction of fuel and product particles with W colloid
different $(\epsilon_{A} > \epsilon_{B})$. This specific choice of
energy parameters is to ensure the motion of W colloid towards T.
%If we choose $\epsilon_{A} >\epsilon_{B}$ for W then A particles will feel more repulsion from W as compare to B particles, which causes W to move towards T. 
To generate the products around colloids an irreversible chemical
reaction $A \rightarrow B$ takes place on the surface of T. Further,
to maintain the system in steady state, the product particles are
converted back to the fuel when they are diffused far from the
colloids.  

In the simulation, all quantities are reported in
dimensionless units. Length, energy, mass and time are
measured in the units of MPC cell length a, $\epsilon$, fluid mass
$m_s$ and $a\sqrt{m_s/\epsilon}$ respectively. Dimension of cubic box
is $L=50$.
%, which is divided into $L^{3}=50^{3}$ cubic cells to perform MPC. 
The MPC rotation angle by which the velocities being rotated
about a randomly chosen axis is fixed at $\phi =120^\circ$. The
collision time step $\tau=0.1$. The average fluid particles density in
a MPC cell $c_{0}$=10. The system temparature is fixed at
$\kappa_{B}T=1$. MD time step $\Delta t=0.01$. The energy parameters
for W interaction are chosen as $\epsilon_{A}=1.0$ , $\epsilon_{B}=0.1$  for fuel and product particles respectively.
%while for $T$ $\epsilon_{A} = \epsilon_{B}=1.0$  
The effective radii of the sphere are $2^{(1/6)} \rho$, where $\rho=2$. Masses of the
colloids are adjusted according to their diameters to ensure density matching with the solvent.

%The transport properties of the fluid depends on the parameters $\tau$, $\phi$ and $N_c$ the
%total number of particles in a cell. Viscosity of the fluid is $\bar{\mu} = m N_c \nu = 7.9$, where $\nu$ is the kinematic viscosity,
%and the diffusion constant for fuel and product particle is
%$D=0.0611$. The Schmidt number is $S_c=\nu/D=13>1$, which ensures that
%momentum transport dominates over mass transport, the Reynolds number
%$R_e= c_0Va/\bar{\mu} < 0.1$, implying that viscosity is dominant over
%inertia, and the P\'{e}clet number $Pe = Va/D <1$, ensures diffusion
%being dominant over fluid advection.

\subsection{Comparison of walker dynamics with power-law biased model}

\begin{figure}[htbp]
	\centering
	\includegraphics[scale=0.33, angle=0]{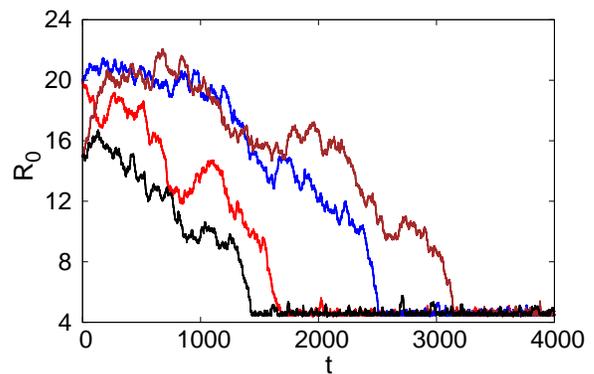}
	\caption{Plot of distance between walker and target as a function of time for different realisations. For two realisations the initial separation $R_0 = 20$, whereas for the others it is $R_0=15$. The time where the distance attains its minimum value is the capture time }
	\label{fig:dis_t}
\end{figure}

The dynamical processes responsible for the biased random walk of the
W involves diffiosiophoresis that imparts directed motion to W and the
thermal fluctuations that randomises its
direction. Fig.~\ref{fig:dis_t} shows some representative trajectories
for the walker depicting the competition between the stochastic and
deterministic components of the processes. The distance-dependent bias towards the target is evident here. 
At large distances, the W sphere exhibits mostly random displacements, whereas when the distance
between W and T goes down, the directed motion dominates leading to eventual capture.

%The dynamical processes that enter this seemingly simple process involve effects that govern the velocity of the W sphere and lead to its eventual capture. 
When the distance between the T and W spheres are large, the
walker encounters with lesser concentration of product $B$ as the
source of the product is situated at the target. This results in the
Brownian motion of the W due to thermal fluctuations.  As the distance
between T and W decreases, the walker experiences enhanced
concentrations of $B$, leading to an increased directed velocity
towards the target. In the microscopic simulations we are able to
probe the concentration fields responsible for the dynamics of W
sphere.

%At large radial distances between the spheres the concentration of product $B$ in the vicinity of W is low and so is its velocity. As the distance decreases the concentration of $B$ increases leading to an increases velocity but as the spheres approach closely more complex interactions lead to the capture event. We are able to probe the details of the mechanism responsible for the capture process through an analysis of the concentration fields that accompany the dynamics.

\begin{figure}[htbp]
	\centering
	\includegraphics[scale=0.33,angle=0]{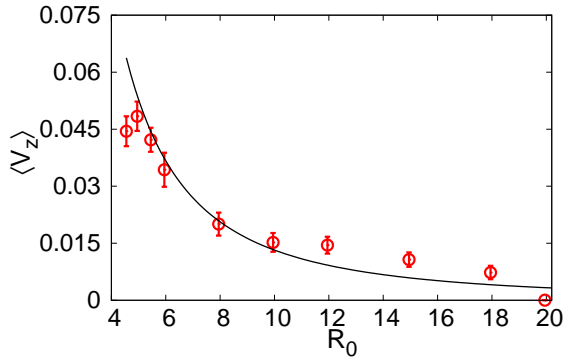}
	\caption{The velocity $V_z$ of walker W as a function of
		the separation $R_{0}$ between the W and T spheres.
		The solid line $R_{0}^{-2} $ fit. Averages were
		obtained from 80 realizations of the dynamics.  }
	\label{fig:vel_r}
\end{figure}

\par	
The directed velocity of the W sphere, $V_z$, is plotted in Fig.~\ref{fig:vel_r} as a function of the distance $R_0$ separating the centers of the two spheres. The figure shows the expected increase in velocity as the W sphere approaches the T sphere.  The velocity profile exhibit a $R_{0}^{-2}$ power law behavior as predicted in the diffusiophoretic theory Eqn.~\ref{eq:vz}. Fig.~\ref{fig:vel_r} shows an excellent agreement between the microscopic simulation and diffusiophoretic theory, except at short distance, where the velocity starts decreasing in simulations. The discrepancy between the microscopic simulations and diffusiophoretic theory at short distance may be due to the slight difference of boundary conditions in these two cases.

 %until, at a short distance, it begins to decrease as the capture event takes place.The velocity profile goes as $R_{0}^{-2}$ power law behavior,but the significant deviations are seen a short distances~\cite{reigh:18dimerform}. This deviation can be explained regarding concentration field.

	\begin{figure}[t]
		\centering
		\includegraphics[scale=0.33,angle=0]{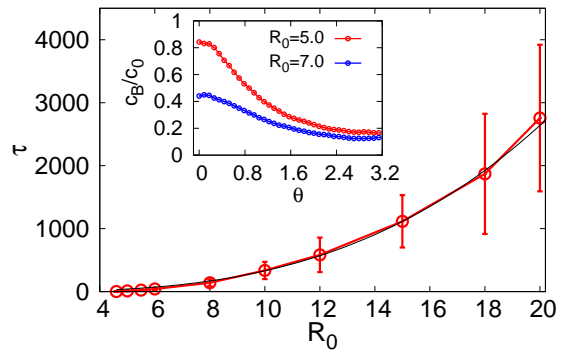}
		\caption{Capture time $\tau$ as a function of the initial separation $R_{0}$ between the spheres.
			The solid line is the $R_{0}^{3}$ power law fit. Inset shows the concentration field of products around W for two different initial distances $R_0=7.0$ and $R_0=10.0$. Simulation results are obtained from averages over 80 realizations.}
		\label{fig:captm_fix}
	\end{figure}

The capture time, $\tau$ which is defined by the time taken by the  W sphere, initially at a distance $R_{0}$ from the target, to reach the T sphere. Fig.~\ref{fig:captm_fix} shows how $\tau$ varies with $R_{0}$. The capture time profile goes as $R_{0}^{3}$ power law both in diffusiophoretic theory and simulation.

%it takes the W sphere, initially at $R_{0}$, to reach the $T$ sphere, i.e., the spheres are separated by a distance equal to the sum of their radii, $R_1+R_2$. .

The concentration and fluid velocity fields vary during the capture process, and these variations play a crucial role in determining the details
of the capture mechanism. The  concentration fields of $B$ species on the surface of the W sphere for two different $R_0$ are shown in the inset
of Fig.~\ref{fig:captm_fix}. It is evident that for the larger
distances between T and W, the gradient is not effective enough for W sphere to respond
whereas for small distances it is prominent. Hence, the bias towards the target is enhanced at shorter distances, giving rise to reduced capture time.

%\begin{figure}[t]
%	\centering
%	%\includegraphics[scale=1.0]{figures/nb_vz_inset.eps}
%	\includegraphics[width=0.48\columnwidth]{diff_conR05.eps}
%	\includegraphics[width=0.48\columnwidth]{diff_conR010.eps}
%	\caption{Normalized tangential gradients $\partial{(c_B/c_0)}/\partial \theta_2$ on the surface of the noncatalytic $S_2$ sphere ($R_2/\sigma=2^{1/6}$) for $R_{0}/\sigma=2.5$ (left column) and
%		$R_{0}/\sigma=5$ (right column), respectively. The angle $\theta$ is the polar angle in spherical polar coordinates where the origin is at the center of the $S_2$ sphere.}
%	\label{fig:diff_fd}
%\end{figure} 

%Fig.~\ref{fig:diff_fd} is normalized tangential gradients for two different separations showing that small separation has a sharp solvent gradient as compare to the large separation between the spheres. 

Hence, the above model for chemotaxis using a pair of colloidal
spheres is a realization of a particular case of the power-law biased
random walker model described earlier, with $\sigma = 2$ in three
dimensions. A quantitative mapping of the results from the two
sections is not part of the scope of the present effort. However, the
strong agreement between the exponents depicted in
Fig.~\ref{fig:tvsr0} and Fig.~\ref{fig:captm_fix} is testimony to the
relevance of the power-law model in the chemotaxis context. Fig.~\ref{fig:tvsr0} corresponds to $\sigma = 0.6$ which 
lies below the critical value (in one dimension), and therefore the mean first passage times are observable within the simulation time scale.
Fig.~\ref{fig:captm_fix}, on the other hand corresponds to $\sigma = 2$, which may well be above the critical value of the power-law model in three
dimension, and yet the timescales are within measurable limits. The reason for this could be that the natural lengthscales and timescales in the two problems
are quite different. As the radius of the walker is shrunk, we found that the capture times became immeasurably large, perhaps indicative that $\sigma = 2$ lies
above the critical value in three dimensions.

The power-law model may prove to be a useful generalization also for
chemotaxis, if an appropriate technique to tweak the exponent may be
engineered. We can envisage applications where it is desirable to bring down the exponent below the critical value so that really tiny capture times may be achieved. One method
by which this could be accomplished is to introduce some type of disorder or crowding~\cite{rajesh2001exact}, which should make the motion of the solute particle sub-diffusive.

\section{Stock market example}

Another potential application of the biased random walk could be in
stock investing. We imagine a scenario where an investor has bought
stock at some time in the past, and finds that its current value is in
a state of loss. The investor would be interested in finding ways to
speed up the first passage of the stock value back to break-even. A
variant model of the power-law biased random walk introduced above may
offer a strategy by which the average time taken till break-even may
be shortened.  The required bias may be applied here `by-hand' by
simply buying more stocks intermittently. In this model, an unbiased
discrete random walk starting from $x_0>0$ makes intermittent jumps
towards the origin. The distance by which the random walk jumps
depends on the position $x$ of the random walk and is proportional to
$\frac{1}{x^\sigma}$ similar to the power-law model. The starting
point of the random walk $x_0$ is equivalent to the net amount by
which a stock price has fallen after it has been bought and the origin
corresponds to the break-even point.

Let $s(t)$ denote the price of a stock at time $t$ and $x(t)$ the
average gain or loss at time $t$. Then $x(t)=s(t)-s_\text{avg}(t)$
where $s_\text{avg}(t)$ is the average amount spent per stock before
time $t$. The position of the random walk $x$ is equivalent to the
negative of $x(t)$. In case of a net loss, stocks are bought
intermittently to reduce the net loss in steps. This action of
reducing the net loss by buying stocks corresponds to the intermittent
jumping of the random walk. The amount by which the net loss is
reduced can be tuned to match the jump size in the model. This tuning
is achieved by deciding the number of stocks that are bought. The
jumping of the random walk or buying of stocks is carried out after
every $t_{\text{watch}}$ steps as long as there is a net loss or until
the random walk crosses the origin.

Suppose after $t_{\text{watch}}$ time steps, at time $t$,
$x(t)<0$. Some number of stocks are bought at the price $s(t)$ so that
$x(t)$ is raised to $x(t)=s(t)-s_\text{avg}(t)$ where
$s_\text{avg}(t)$ is now the average amount spent per stock after
buying stocks costing
$s(t)$. Since we had $s(t)<s_\text{avg}(t)$, after buying stocks at $s(t)$, $s_\text{avg}(t)$ will have reduced.

If the jump size of the random walk is fixed by a constant $\gamma$ so
that the jump length
$\frac{\gamma}{x^\sigma}\equiv\frac{\gamma}{|x(t)|^{\sigma}}$, the
number of stocks to be bought after every $t_{\text{watch}}$ time
steps can be found. Suppose at $t=t_0$, $N_0$ stocks were bought and
at $t=t_1$, there was a huge net loss and the process of buying stocks
intermittently was started. If $N_1$ stocks were bought at $t_1$,
$N_2$ stocks at $t_2=t_1+t_{\text{watch}}$ and similarly $N_{i-1}$
stocks were bought at $t_{i-1}=t_1+(i-2)t_{\text{watch}}$, we have
\begin{equation}
    s_\text{avg}(t_i)= \frac{N_0s(t_0)+N_1s(t_1)+\cdots+N_{i-1}s(t_{i-1})}{N_0+N_1+\cdots+N_{i-1}}
\end{equation}
for the net amount spent per stock after buying stocks at $t=t_{i-1}$. The net loss has been reduced. In terms of the number of stocks bought,

\begin{equation}
\begin{aligned}
    \frac{N_0s(t_0)+\cdots+N_{i-1}s(t_{i-1})}{N_0+\cdots+N_{i-1}}-\frac{N_0s(t_0)+\cdots+N_{i}s(t_{i})}{N_0+\cdots+N_i}\\=\frac{\gamma}{|x(t_i)|^{\sigma}}.
\end{aligned}
\label{eq:r}
\end{equation}

Upon rearranging the left-hand side of Eqn.~\ref{eq:r}, we see that the number of stocks to be bought at $t=t_i$ is
\begin{equation}
    N_i=(N_0+N_1+\cdots+N_{i-1})\frac{\gamma}{-x(t_i)|x(t_i)|^{\sigma}-\gamma}.
\end{equation}

\begin{figure}
\centering
\includegraphics[scale=0.33]{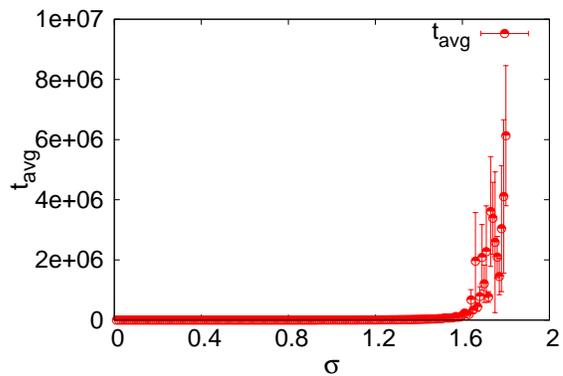}
\caption{Mean first-passage time to the origin $t_{\text{avg}}$ for different values of $\sigma$ with $\gamma=600$, $t_{\text{watch}}=10$ and $x_0=500$. The high error-bars after $\sigma\sim1.6$ is indicative of a transition from finite to infinite mean first-passage time.}
\label{fig:n}
\end{figure}

In the random walk model, $x(t_1)$ corresponds to the initial position
$x_0$ of the unbiased random walk. The random walk makes jumps of
length $\frac{\gamma}{x^{\sigma}}$ where $x\equiv|x(t)|$ at $t=t_1$,
$t=t_2$ and so on until the random walk reaches the origin. The mean
first-passage time to the origin is found for different values of
$\sigma$ and a critical value of $\sigma$ is seen beyond which the
mean first-passage time becomes arbitrarily large. This can be seen in
Fig.~\ref{fig:n} for a random walk starting from $x(0)=500$ making
jumps towards the origin with $t_{\text{watch}}=10$ and
$\gamma=600$. The large error bars seen for higher values of $\sigma$
indicate infinite mean first-passage time to the origin. From the
plot, the critical value of $\sigma$ seems to be around
$\sigma_c\approx1.6$.
%\newline

The constants $\gamma$ and $t_\text{watch}$ can be chosen in a manner
to optimize the time taken to break even and to minimize the total
amount that needs to be invested. The value of $\sigma$ should be chosen
to lie below $\sigma_c$ so that the average time required to break even is expected to be finite. Of course this model relies on the assumption that
the stock is performing an unbiased random walk, and the only bias comes from the investor, which is in general not true. However, the strength of this
strategy lies in the identification of a critical value of the parameter $\sigma$ below which the sucess potential of the strategy is greatly enhanced.

\section{Concluding Remarks}
We have introduced a one-parameter distance-dependent power-law biased
random walk model in which the first passage properties show
dramatically different behavior depending on the value of the
parameter $\sigma$. In one dimension, when the value of $\sigma$ is
below a critical value $\sigma_c \approx 1.14$, we report a finite
value of the mean first-passage time. For $\sigma > \sigma_c$, the
motion is similar to that of the unbiased random walk, where although
the probability of return is unity, the mean first-passage time is
infinite. This is understood from the first-passage probability which
is posited to go as $f_{n,m} \sim\frac{1}{n^{\alpha}}$, where $\alpha$
itself depends on $\sigma$. When $\sigma <\sigma_c$, $\alpha$ is
greater than $2$, thus making the summation $\sum nf_{n,m}$
convergent, which in turn yields a finite mean first-passage
time. With the help of some novel numerical techniques, we show how an
exact simulation of these biased random walks becomes possible, if one
restricts the study to a finite maximum time. The value of the maximum
time is only limited by memory resources.

We point out that this model is closely connected to the real-world problem
of chemotaxis, where distance-dependent bias arises naturally.
The phenomenon of chemotaxis is well described using diffusiophoretic theory, supplemented with
microscopic simulations to obtain first passage properties.
It turns out that chemotaxis corresponds to the three-dimensional version
of the above power-law biased random walk model with the specific parameter value of $\sigma = 2$,
due to the inherent diffusive nature of the solvent transport. Despite 
$\sigma$ being apparently larger than the critical value, finite capture times
are seen in simulations. The analogy between the power-law model and chemotaxis model should be
viewed with caution, because in the latter, the walker and the target are not point-particles. The finite-radii 
of the walker and the target ensure that a weaker bias is still effective in directed motion of the walker, and hence
we see finite capture times even when $\sigma$ is apparently greater than the critical value. We have found that
reducing the radii of the spheres leads to larger capture times, which after a point become inaccessible within 
the simulations. 

The second example to which we attempt to apply our model is to the
problem of a stock investor devising a strategy to extricate himself
from loss. Here, we introduce a variant of the power-law biased random
walk model, where on top of an unbiased random walk, we embed
intermittent jumps directed towards the target, whose size is determined by the power-law. Similar
to the above model, there is a critical $\sigma_c \sim 1.6$ below
which mean first passage times are finite. More tests based on real
data from stock portfolios would be desirable to explore whether this
strategy may be adopted in a real-world scenario. This stock market
problem is inherently one-dimensional and therefore the model is
well-suited for such tests.

We anticipate that the proposed power-law biased random walk model may find application in other phenomena. It would be particularly
interesting to consider higher dimensional generalizations of this model, although memory requirements would make the simulations more challenging. 
Incorporation of crowding and disorder effects into the chemotaxis scenario might provide a knob which can tune the underlying $\sigma$ in
the model. An improvised power-law model with finite radii of the walker and target might provide an alternative route to make more quantitative 
connections to the chemotaxis scenario. It would be exciting if this model can provide fresh insights into stock investing strategy.\\

\acknowledgements AS is grateful to Kavita Jain for stimulating discussions and acknowledges support from the DST-INSPIRE Faculty Award [DST/INSPIRE/04/2014/002461]. 
Some of the simulations in this project were run on the High Performance Computing (HPC) facility at IISER Bhopal. CP is grateful to Sangram Kadam for help with the computational work. 

 \bibliography{refs}

\end{document}